\documentclass[10pt]{iopart}
\usepackage{iopams}
\expandafter\let\csname equation*\endcsname\relax
\expandafter\let\csname endequation*\endcsname\relax
\usepackage{amsmath}
\usepackage{graphicx}
\usepackage{bm,epsfig}
\usepackage{mathrsfs}
\usepackage{color}
\usepackage{xcolor}
\usepackage{color,soul}

\usepackage{cite}
\usepackage{gensymb}
\allowdisplaybreaks

\begin{document}
\title[Azimuthal modulation of EIT by using aLG beams]{Azimuthal modulation of electromagnetically-induced transparency by using asymmetrical Laguerre-Gaussian beams}

\author{Seyedeh Hamideh Kazemi \& Mohammad Mahmoudi}

\address{Department of Physics, University of Zanjan, University Blvd., 45371-38791, Zanjan, Iran}
\ead{mahmoudi@znu.ac.ir}
\vspace{10pt}
\begin{indented}
\item[]August 2019
\end{indented}

\begin{abstract}
Recently, the generation and detection of structured light field have drawn a great deal of attention, due
to their importance in high-capacity optical data storage and quantum technology. In this letter, we explore the azimuthal modulation of optical transparency in a four-level double-V type quantum system near a plasmonic nanostructure. A Laguerre-Gaussian beam and the interaction of the system with free-space vacuum modes have been employed to create the phase-dependent absorption of a non-vortex probe field. First, we demonstrate how to identify the azimuthal index associated with the conventional LG beam via measuring the probe absorption so that the phase information of such a beam gets encoded on the spatially-dependent absorption profile with angularly-distributed lobes. Also, a spatially-varying
optical transparency can be formed, due to the periodic variation of the absorption spectrum. Then, asymmetrical Laguerre-Gaussian beams are used to extend the selective spatial transparency mechanism to
asymmetric spatially-structured windows, allowing for optical manipulation of spatial modes at arbitrary positions. Moreover, we 
investigate the influence of the asymmetric parameter on the features of the spatial inhomogeneities and show how the beams enable us to imprint the phase information of the orbital angular momentum at a desired position.

\end{abstract}
%around 198 words

\pacs{42.50.Gy, 78.67.Bf, 73.20.Mf, 42.50.Tx}
%42.50.Gy Effects of atomic coherence on propagation, absorption, and amplification of light; electromagnetically induced transparency and absorption
%78.67.Bf Nanocrystals, nanoparticles, and nanoclusters
%73.20.Mf – Collective excitations (including excitons, polarons, plasmons and other charge-density excitations)
%42.50.Tx Optical angular momentum and its quantum aspects

\vspace{2pc}
\noindent{\it Keywords}: azimuthal modulation, probe absorption, plasmonic nanostructure, asymmetric Laguerre-Gaussian beam

%\submitto{\LPL}
\ioptwocol

\section{Introduction}
The coherent interaction of light with atoms has attracted much attention, due to its applications for quantum optics and quantum information. Perhaps the most prominent and well-known example of such an interaction is electromagnetically-induced transparency (EIT) in which the destructive quantum interference between probability amplitudes of two transition pathways results in the elimination of absorption along with a deceleration of the group velocity at the transparent point \cite{Harris}. This phenomenon leads to a variety of practical applications, for instance, highly enhanced nonlinear susceptibility in the spectral region of induced transparency accompanied with steep dispersion \cite{Boller}, slow and stopped light \cite{Kasapi,Hau}, multi-wave mixing \cite{Saldana,Wang3}, optical soliton \cite{Deng} and so on. Also, techniques based on EIT allow for the storage of optical data in matter, facilitating the realization of all-optical quantum information processing and quantum memory \cite{Fleischhauer,Lvovsky}. On the other hand, recent studies have indicated that coherent optical phenomena of quantum systems such as atoms, molecules and quantum dots can be strongly influenced in the presence of nanostructures \cite{Klimov,Mori}. A prominent feature of such systems is remarkably modified spontaneous decay rate in different dipole moment directions. For instance, Yannopapas \textit{et al.} showed how placement of a three-level quantum emitter in
the proximity of plasmonic nanostructures such as metallic slabs, nanospheres, or periodic arrays of metal-coated spheres can increase the degree of quantum interference in a spontaneous emission \cite{Yannopapas}.

Optical vortex beam carrying orbital angular momentum (OAM) have attracted great interest owing to numerous applications ranging from optical imaging, telecommunications, and optical tweezers to quantum technology \cite{Padgett}. The OAM of light, with inherent infinite dimension, gives additional possibilities in manipulation of the optical information in high-dimensional Hilbert spaces, providing higher efficiency and enhanced capacity. It was Allen \textit{et al.} who recognized that beams with the phase cross-section of the form $\exp(i n \varphi)$ carry an OAM equivalent to $n \hbar$ per photon (with $n$ as an integer and $\varphi$ being as the azimuthal coordinate) \cite{Allen}. An example is Laguerre-Gaussian (LG) mode, exhibiting a doughnut-type intensity profile and a helical phase structure. Such modes have found use in areas such as quantum and nano-optics \cite{8,Bhattacharya,das,9,10,Sabegh,local}, communications and quantum information processing \cite{15,16,17}. In the domain of optical micro-manipulation, LG modes have also attracted much interest, setting microscopic particles into rotation, trapping low-refractive index particles and creating micropumps \cite{3,4,5}. Recently, a family of asymmetric Laguerre-Gaussian (aLG) laser beams, with a crescent-shaped intensity pattern and a nonlinear OAM, is introduced by Kovalev and his co-workers \cite{Kovalev}. A rotation of large asymmetrical nanoparticles with increased velocity by using aLG beams has been carried out in another work of the authors \cite{Kovalev2}. More recently, the microscopic interaction of aLG beams with ultra-cold atoms has been investigated, resulting in a superposition of matter vortex states in an atomic Bose-Einstein condensate \cite{Subrata}.  However, to the best our knowledge, no study to date has explored  the modulation of transparency by using aLG beams.

The interaction between atomic systems and optical vortex beams is a major focus of scientific interest and 
has been exploited in context of EIT \cite{Pugatch,16,Han,Radwell,Sharma1,Hamedi,Sharma} and four-wave mixing \cite{Marino,Walker,cao}. For example, the storage of optical vortex, based on an EIT protocol, has been reported in atomic ensembles \cite{Pugatch,16}. 
Several studies have also demonstrated how the properties of EIT can be utilized to identify the OAM information of the vortex light. In the seminal work by Radwell \textit{et al.}, an atomic scheme for creating structured beams was suggested, by using a probe beam with azimuthally-varying phase and polarization structure, so that the self-modulation
of incident light beams can convert the phase into intensity information \cite{Radwell}. More recently, another scheme has been proposed for the detection of structured light fields via measuring the absorption profile, which takes advantage of a closed-loop structure of a highly resonant atom-light coupling. Here, we suggest another method to detect the characteristics of the OAM carried by a LG field, and spatially-varying optical transparency via measuring the absorption of a non-vortex probe beam. 

In this paper, through combining the phase-dependent optical effects and OAM, we propose a four-level double-V
type quantum system for realizing azimuthal modulation of EIT. By exposing the system to a non-vortex probe field and a
field carrying OAM, we find that the absorption of the non-vortex probe field depends on the azimuthal factor of the vortex LG beam. First, we consider the scheme in which the quantum system is exposed to a conventional LG beam and show how the phase-dependent profile reveals both the magnitude and sign of the azimuthal index associated with the LG field. Moreover, one can obtain information about positions of low light transmission, gain or optical transparency through measuring the absorption profile. Then, we demonstrate
how aLG beams affect the features of spatial inhomogeneities and how the beams enable us to imprint the phase
information of OAM onto the probe absorption at specific positions which can be controlled easily by the shifts.

%This modulations enable us to imprint phase information .
%One can identify different regions of spatially structured transparency through measuring the absorption of probe field.
%%%%%%%%%%%%%%%%%%%%%
%the probe absorption can be reflected by a standard spectroscopic method, whch may be realized via the experiment proposed in Ref. \cite{Radwell}.
%%%%%%%%%%%%%%%%%%%%%%%%%%%%%%%%%%%%%%%%%%%%%%%
%As mentioned previously, the phase factor associated with LG beam [$\exp(i n \varphi)$] in equation~(\ref{eq5}) gives rise to spatial inhomogeneity of the medium; 
%bright area related to the OAM index associated with LG field.
%interestingly, the two spots are suppressed. 
%%%%%%%%%%%%%%%%%%%%%%%%%%%
%The suggested scheme can simplify a possible implementation of high-capacity data storage technologies and quantuminformation.
%the presence of an LG beam carrying an optical vortex leads to azimuthal variation of the dynamics of the system so that

\section{Asymmetric Laguerre-Gaussian beams}

LG beam, an eigenmode of the paraxial propagation electromagnetic equations, exhibits orbital angular momentum of $n\hbar$ per photon, because of its helical wavefront and on-axis phase singularity \cite{Allen}. The laser mode is usually denoted by LG$_{m}^{n}$, with $n$ and $m$ being as azimuthal and radial indices. The former is the number of times the phase completes 2$\pi$ on a closed-loop around the axis of propagation and $m+1$ denotes the number of radial nodes in the intensity distribution \cite{hanle}. Denoting coordinates as ($x$,$y$,$z$)=($r,\varphi$,$z$) in either Cartesian or cylindrical, the complex amplitude of the conventional LG-mode in the initial plane is written as  
\begin{align}
E (r,\varphi,z=0)&= E^{'} \,(\dfrac{ \sqrt{2} \,r}{\mathrm{w}})^{\vert n \vert} \, L_{m}^{n}(\dfrac{ 2 r^{2}}{\mathrm{w}^{2}})\\ \nonumber
&\exp ( - \dfrac{r^2}{\mathrm{w}^{2}} +i n \varphi ) +\mathrm{c.c.},
\label{equ1}
\end{align}
where, we have defined $\exp(i n \varphi)$ as the azimuthal phase factor, with $\varphi=\arctan (y/x)$. Also, the parameter $\mathrm{w}$ is the beam waist and $L_{m}^{n}(x)$ is the associated Laguerre polynomial. The customary Gaussian mode can be viewed as an LG mode with zero indices (LG$_{0}^{0}$), having a pattern of a round spot with a constant phase \cite{Steiner}. The profile of LG modes show concentric rings of intensity, the number of which is determined by the radial index $m$, and modes with increasing $n$ show angularly-distributed lobes so that there are 2$n$($m$ + 1) spots in their transverse mode patterns \cite{gotte}. The handedness of helical wave fronts of these modes is related to the sign of azimuthal index ($n$) and can be chosen by convention. It is well known that beams with such a phase factor of $\exp(i n \varphi)$ have a well-defined OAM, which can be linked to the phase factor of an optical vortex (\textit{i.e}., a phase singularity of an optical field) so that the beams are considered to have orbital vortices with a topological charge of $n$ \cite{dennis}. 

In the Cartesian coordinates, the OAM term in the equation~(1), \textit{i.e}., $r^n\, \exp(i n \varphi)$, looks like ($x$+$iy$)$^{n}$; if the beam is shifted by the vector ($\delta x$,$\delta y$)-along the coordinates $x$ and $y$, respectively- the beam amplitude in Cartesian coordinates takes the form of
\begin{align}
E (x,y,z=0)&= E^{''} \,(\dfrac{ \sqrt{2}}{\mathrm{w}})^{\vert n \vert} \, [(x-\delta x)+i(y-\delta y)]^{n} \\  \nonumber
&L_{m}^{n}(\dfrac{ 2 \rho^{2}}{\mathrm{w}^{2}})\, \exp ( - \dfrac{\rho^2}{\mathrm{w}^{2}} ) +\mathrm{c.c.},
\label{eq10}
\end{align}
where, $\delta x$ and $\delta y$ can take complex values and $\rho$=$\sqrt{(x-\delta x)^2+(y-\delta y)^2}$. As in conventional LG beams, the transverse intensity pattern of aLG modes consists of a finite number of rings ($m$+1), but with a nonuniform distribution. Generally speaking, these LG beams, without modal properties, show an asymmetric intensity pattern in a plane perpendicular to the propagation axis ($z$-axis); shifts can cause either displacement or distortion (or even both of them) so that in contrast to conventional LG beams, the intensity pattern of shifted beams does not preserve axial symmetry and the degree of asymmetry can be determined by the parameters $\delta x$ and $\delta y$. If shifts in transverse coordinates ($x$ and $y$) are real, the beam is just displaced and not deformed. While, the shape of a shifted LG beam is distorted for the case of complex-valued shift(s); if both shifts are imaginary, the shape of the beam is distorted. For the case of a real shift along one coordinate and an imaginary-valued shift along the other, the beam is displaced and its shape is distorted, as well \cite{Kovalev}. In this paper, we consider three different cases for calculations concerning aLG beams: 1) shift(s) is real; 2) one shift is real along one coordinate and imaginary along other ($\delta x$=$i \delta y$=$i a \mathrm{w}$), and 3) both shifts are purely imaginary ($\delta x$=$\delta y$=$i a \mathrm{w}$), with $a$ as a real number (hereafter, we call it "asymmetry parameter").

Throughout the paper, for simplicity, we have assumed $m$ = $0$ allowing only for the so-called doughnut modes of order $n$, however, the treatment can be applied also to LG beams with non-zero radial index. Rabi frequency for the conventional LG beams, without any off-axis radial node, is written as  
\begin{equation}
g (r,\varphi)=  g^{'} \,(\dfrac{ \sqrt{2} \,r}{\mathrm{w}})^{\vert n \vert} \,\exp ( - \dfrac{r^2}{\mathrm{w}^{2}} +i n \varphi ),
\end{equation}
with $g^{'}$=$(\mu \,E^{'})/{\hbar}$ being as Rabi frequency constant of the conventional LG beams and $\mu$ as the transition dipole moment. Similarly, the expression for Rabi frequency of the shifted beams with $m$=0 takes the form of $g^{''} \,(\sqrt{2}/\mathrm{w})^{\vert n \vert} \, [(x-\delta x)+i(y-\delta y)]^{n}$ $\, \exp ( - \rho^{2}/\mathrm{w}^{2} )$, with $g^{''}$=$(\mu \,E^{''})/{\hbar}$.
%\begin{align}
%g_{ij} (x,y)&=  g^{'}_{ij} \,(\dfrac{ \sqrt{2}}{w})^{n}  \, [(x-\delta x)+i(y-\delta y)]^{n} \\  \nonumber
%&L_{m}^{n}(\dfrac{ 2 \rho^{2}}{w^{2}})\, \exp ( - \dfrac{\rho^2}{w^{2}} ).
%\end{align}

\begin{figure*}[!ht]
\centerline{\includegraphics[width=0.7\linewidth]{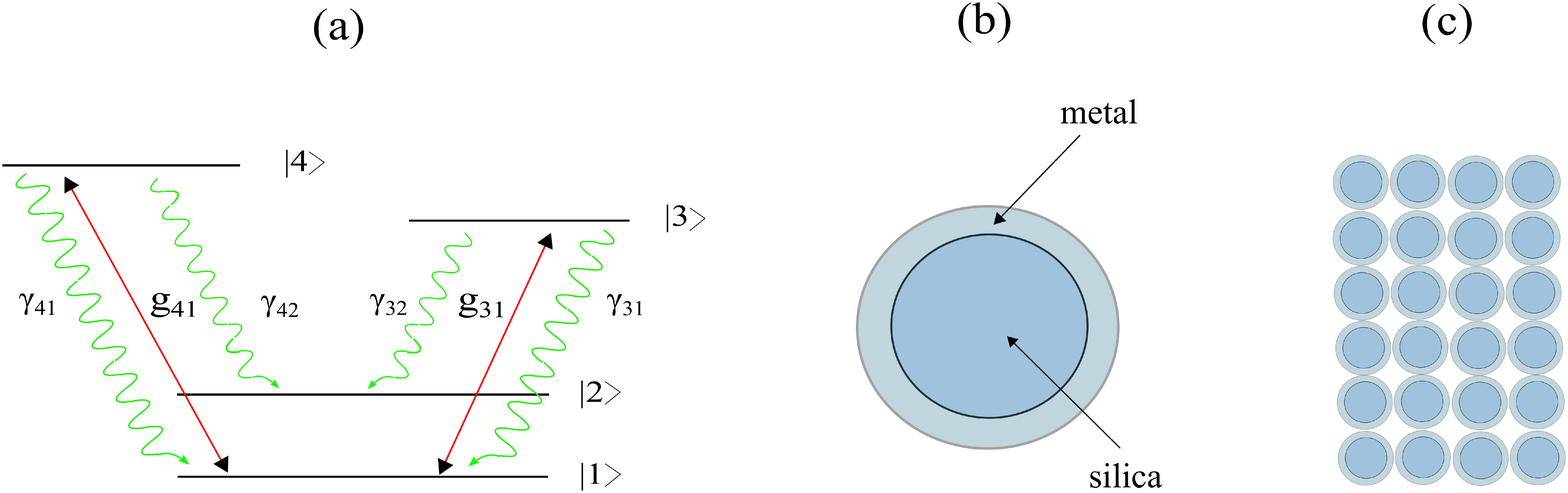}}
\caption{ (a) The energy-level structure of the four-level double-$V$ type quantum system, interacting with two circularly polarized weak probe laser fields $g_{31}$ and $g_{41}$. The decay rate of the corresponding levels is denoted by $\gamma_{ij}$($i,j \in \lbrace 1,...,4 \rbrace$ and $i>j$). (b) A metallic nanosphere made from a silica core and metal coating and (c) a two-dimensional array of such nanospheres. }
\label{fig1}
\end{figure*} 

\section{Atom near the plasmonic nanostructure}

Our analysis is based on the scheme depicted in figure~\ref{fig1}; a double-$V$ type atomic system with two Zeeman
sublevels for upper states $\vert 3 \rangle$ and $\vert 4 \rangle$, and two lower ones, $\vert 1 \rangle$ and $\vert 2 \rangle$. The transitions $\vert 3 \rangle \rightarrow \vert 1 \rangle$ and $\vert 4 \rangle \rightarrow \vert 1 \rangle$ are coherently driven by two laser fields, with Rabi frequencies $g_{31}$ and $g_{41}$, respectively. Also, the decay rate from the upper states to the lower ones is denoted by 2$\gamma_{ij}$ ($i,j$ $\in$ $\lbrace 1,...,4 \rbrace$ and $i>j$). The expression for the dipole-moment operator in the $x$-$z$ plane is given by $\vec{\mu}$ = $\mu^{'} (\vert 3 \rangle \langle 1 \vert \hat{\varepsilon}_{-} + \vert 4 \rangle \langle 1 \vert \hat{\varepsilon}_{+}  )$ + $\mu^{''} (\vert 3 \rangle \langle 2 \vert \hat{\varepsilon}_{-} + \vert 4 \rangle \langle 2 \vert \hat{\varepsilon}_{+}  )$ + $\mathrm{H.c.}$, in which $\hat{\varepsilon}_{+}$=$(\mathrm{\textbf{e}}_z + i \mathrm{\textbf{e}}_x)/\sqrt{2}$ and $\hat{\varepsilon}_{-}$=$(\mathrm{\textbf{e}}_z - i \mathrm{\textbf{e}}_x)/\sqrt{2}$
refer to right-rotating and left-rotating unit vectors, respectively. Also, H.c. corresponds to the Hermitian conjugate of the terms explicitly written in the relation. Noting that the amplitude of the dipole moments, \textit{i.e}., $\mu^{'}$ and $\mu^{''}$ are taken to be real. 

Two circularly polarized continuous-wave laser fields whose electric field form is
$\vec{E} = \hat{\varepsilon}_{+} E_{31} \cos(\varpi_{31} t + \phi_{31})$ + $ \hat{\varepsilon}_{-} E_{41} \cos(\varpi_{41} t + \phi_{41})$ are interacting with the quantum system. Here, $E_{31}(E_{41})$, $\varpi_{31}(\varpi_{41})$ and $\phi_{31}(\phi_{41})$ denote the electric-field amplitude, angular frequency and the phase of field, respectively. In the following, we consider that the system is located in vacuum at a distance $d$ away from the surface of the plasmonic nanostructure in the $x$-$y$ plane. As is seen in figures~\ref{fig1}(b) and \ref{fig1}(c), the nanostructure is a two-dimensional array of touching metal-coated silica nanospheres. It may be noted that such a scheme has been already studied for investigating phase-dependent optical effects in a four-level quantum system near a plasmonic nanostructure \cite{Paspalakis}. 

Then, the Hamiltonian of the system can be written as
\begin{align}
H&=  \, \hbar \,(-\delta - \dfrac{\omega_{43}}{2}) \vert 3 \rangle \langle 3 \vert + \hbar \,(-\delta + \dfrac{\omega_{43}}{2}) \vert 4 \rangle \langle 4 \vert \\  \nonumber
&- (\dfrac{\hbar g_{31} e^{i \phi_{31}}  }{2}  \vert 1 \rangle \langle 3 \vert +   \dfrac{\hbar g_{41} e^{i \phi_{41}}  }{2} \vert 1 \rangle \langle 4 \vert +\mathrm{H.c.}),
\label{eq1}
\end{align}
where $\delta$=$\omega - \bar{\omega}$ denotes the detuning from resonance with the average transition energies of two upper states from the state $\vert 1 \rangle$, with $\bar{\omega} = (\omega_{3} + \omega_{4})/2 -\omega_{1}$, $\omega_{43}= (\omega_{4} - \omega_{3}) /2$ and $\omega_i$ being as the energy of state $\vert i \rangle$. Also, the Rabi frequencies are defined as $g_{41}= \mu^{'} E_{41} /{\hbar} $ and $g_{31}= \mu^{'} E_{31} /{\hbar} $. 

Now, we assume that the transitions from $\vert 3 \rangle$ and $\vert 4 \rangle$ to $\vert 2 \rangle$ lie
within the surface plasmon bands of the nanostructure, while both transitions from $\vert 3 \rangle$ and $\vert 4 \rangle$ to $\vert 1 \rangle$ are  spectrally distinct from the surface-plasmon bands, and subsequently, the spontaneous decay in the latter transitions ($\gamma_{31}$ and $\gamma_{41}$) occurs due to the interaction of the system with free-space vacuum modes (after that, we call them "free-space spontaneous decay rates"). The master equation describing the dynamical evolution of the system can be written as $d\rho/dt$=$ -i/\hbar \, [H,\rho] + \mathcal{L}[\rho]$. According to Ref. \cite{Paspalakis}, the equations for the density matrix elements of the system are 
\begin{subequations}
\begin{align}
\dot{\rho}_{11}&= 2 \gamma^{'} (\rho_{33} + \rho_{44}) -  \dfrac{i}{2} (\rho_{13} \,g_{31}^{*}  e^{-i\phi_{31}} - \rho_{31} \,g_{31} e^{i\phi_{31}}  ) \\   \nonumber
&-   \dfrac{i}{2} (\rho_{14} \, g_{41}^{*} e^{-i\phi_{41}} - \rho_{41} \, g_{41} e^{i\phi_{41}}  ),\\ 
\dot{\rho}_{33}&= -2 (\gamma^{'} + \gamma)  \, \rho_{33} +  \dfrac{i}{2} (\rho_{13} \, g_{31}^{*} e^{-i\phi_{31}} - \rho_{31} \, g_{31}  e^{i\phi_{31}}  ) \\ \nonumber  
&-  \kappa (\rho_{34} + \rho_{43}),\\
\dot{\rho}_{44}&= -2 (\gamma^{'} + \gamma) \, \rho_{44} + \dfrac{i}{2} (\rho_{14} \, g_{41}^{*} e^{-i\phi_{41}} - \rho_{41} \,g_{41} e^{i\phi_{41}}  ) \\ \nonumber  
&-  \kappa (\rho_{34} + \rho_{43}),\\
\dot{\rho}_{31}&= [i \delta + i \dfrac{\omega_{43}}{2}  - (\gamma + \gamma^{'})] \, \rho_{31} +  \dfrac{i}{2}  g_{31}^{*}  e^{-i\phi_{31}} (\rho_{11}  - \rho_{33}) \\ \nonumber  
&-   \dfrac{i}{2} g_{41}^{*} e^{-i\phi_{41}} \, \rho_{34} - \kappa \, \rho_{41} ,\\
\dot{\rho}_{41}&= [i \delta - i \dfrac{\omega_{43}}{2}  - (\gamma + \gamma^{'})] \, \rho_{41} +  \dfrac{i}{2} g_{41}^{*} e^{-i\phi_{41}} (\rho_{11}  - \rho_{44}) \\ \nonumber  
&-   \dfrac{i}{2} g_{31}^{*} e^{-i\phi_{31}} \, \rho_{43} - \kappa \, \rho_{31} ,\\
\dot{\rho}_{34}&= [i \omega_{43}   - 2 (\gamma+\gamma^{'}) ] \, \rho_{34} +  \dfrac{i}{2}  g_{31}^{*} e^{-i\phi_{31}} \, \rho_{14}   \\ \nonumber  
&-   \dfrac{i}{2} g_{41}  e^{i\phi_{41}} \, \rho_{31} - \kappa (\rho_{33} + \rho_{44} ).
\end{align}
\label{equation2}
\end{subequations}
The remaining equations follow from $\rho_{ij}$=$\rho^{*}_{ji}$ and trace condition $\sum _{i} \rho_{ii}$=1. Noting that we choose the energy difference of two Zeeman upper states to be small, \textit{i.e}., $\omega_{43}$ to be just a few $\Gamma_{0}$ (that is the decay rate of states $\vert 3\rangle$ and $\vert 4\rangle$ to state $\vert 2\rangle$ in the vacuum), therefore, the approximations $\gamma_{41}$=$\gamma_{31}$=$\gamma^{'}$ and $\gamma_{42}=\gamma_{32}=\gamma$ would be reasonable. Also, $\kappa$ denotes the coupling coefficient between two upper states, due to spontaneous emission in a modified anisotropic vacuum \cite{28,29}. The coefficient, which is responsible for the quantum interference between two decay channels, can be evaluated from the relation
\begin{align}
\kappa&= \dfrac{\mu_{0} \mu^{2} \tilde{\omega}^2 }{ 2 \hbar} \mathrm{Im}[ G_{zz} (r,r;\tilde{\omega}) - G_{xx} (r,r;\tilde{\omega})] \\  \nonumber
&= \dfrac{1}{2} (\Gamma_{\bot} - \Gamma_{\|}),
\label{eq3}
\end{align}
in which $G_{ij}(r,r;\tilde{\omega})$ with $i,j$=$x,y,z$ are the components of the Green’s tensor and $r$ refers to the
position of the quantum emitter. We further define $\Gamma_{\bot}$=$\mu_{0} \mu^{2} \tilde{\omega}^2 \mathrm{Im}[G_{zz}(r,r;\tilde{\omega})]/ \hbar $ as the spontaneous decay rate of the dipole momentum perpendicular to the interface (along the $z$-axis), and $\Gamma_{\|}$=$\mu_{0} \mu^{2} \tilde{\omega}^2 \mathrm{Im}[G_{xx}(r,r;\tilde{\omega})]/ \hbar $ as that parallel to the interface (along the $x$-axis). In addition, $\mu_{0}$ 
is the permeability of vacuum and $\tilde{\omega}=(\omega_{3} + \omega_{4})/2 -\omega_{2}$.
 
Also, the free-space spontaneous decay rate from two upper states to state $\vert 2\rangle$, due to the interaction of
the quantum system with surface plasmon bands of the plasmonic nanostructure, is expressed as
\begin{align}
\gamma&= \dfrac{\mu_{0} \mu^{2} \tilde{\omega}^2 }{ 2 \hbar} \mathrm{Im}[ G_{zz} (r,r;\tilde{\omega}) + G_{xx} (r,r;\tilde{\omega})] \\  \nonumber
&= \dfrac{1}{2} (\Gamma_{\bot} + \Gamma_{\|}).
\label{eq4}
\end{align}
What is also worth mentioning is that we adopt the ratio $p=(\Gamma_{\bot} - \Gamma_{\|})/(\Gamma_{\bot}+ \Gamma_{\|})$ to measure quantum interference. It is clear that the degree of the quantum interference ($p$) increases with the difference between $\Gamma_{\bot}$ and $\Gamma_{\|}$; the larger difference is, the stronger the quantum interference. For $\gamma$=$\kappa$ (or $\Gamma_{\|}$=0), we have $p$=1, the maximum quantum interference in spontaneous emission, which can be achieved by placing the emitter close to a structure that completely quenches $\Gamma_{\|}$ \cite{Paspalakis,29}. On the other hand, the degree of the quantum interference would be zero, if we place the emitter in vacuum: $\kappa=0$ or $\Gamma_{\|}=\Gamma_{\bot}$.

On the other hand, the dispersion and absorption for a weak probe laser field, is determined by the linear susceptibility of the weak probe field, which can be written as $\chi$=$(2 N \mu^{'} \rho_{31})/(\epsilon_0 E_{31} \, e^{-i\phi_{31}})$ = $( N \mu^{'2} e^{i\phi_{31}}\, \rho_{31}) /(\hbar \epsilon_0  \, g_{31})$ with $N$ being as density of the quantum system. By setting $2 N \mu^{'2} / \hbar \epsilon_0$=1, for simplicity, the susceptibility now reads $\chi$ = $ \e^{i \phi_{31}} \rho_{31}/ g_{31}$. Using perturbation theory and assuming that both fields are weak, we can calculate the steady-state solution for $\rho_{31}$ from equations~(\ref{equation2}) and the expression for the medium susceptibility can be written as \cite{Paspalakis}
\begin{equation}
\chi= \dfrac{A}{B} -\dfrac{ i \kappa \, g_{41} \, e^{i  \phi}}{  B \,g_{31}},
\label{eq5}
\end{equation}
where
\begin{subequations}
\begin{align}
A&= \delta-\dfrac{\omega_{43}}{2} + i\,(\gamma + \gamma^{'}), \\
B&=(   -i \delta + i \dfrac{\omega_{43}}{2}  +\gamma + \gamma^{'} )\,( -i \delta - i \dfrac{\omega_{43}}{2} +\gamma + \gamma^{'}  ) - \kappa^{2}.
\end{align}
\end{subequations}
Here, $\phi$=$\phi_{31}$-$\phi_{41}$ denotes the relative phase of applied fields. Throughout the paper, we consider the case in which the field $E_{41}$ has an LG profile with $m$=0, either a conventional or an aLG one, but the other weak field has no vortices. Therefore, the susceptibility, and subsequently, the linear absorption of the probe field depend on the OAM and the azimuthal angle of the field ($\varphi$), as the field carries an optical vortex. In addition, the above analytic expression shows the dependence of the medium susceptibility on the relative phase of applied fields ($\phi$) so that the parameter can also play a decisive role in the modulation of the absorption.

\section{Results and Discussions}

\begin{figure*}[!t]
\centerline{\includegraphics[width=0.9\linewidth]{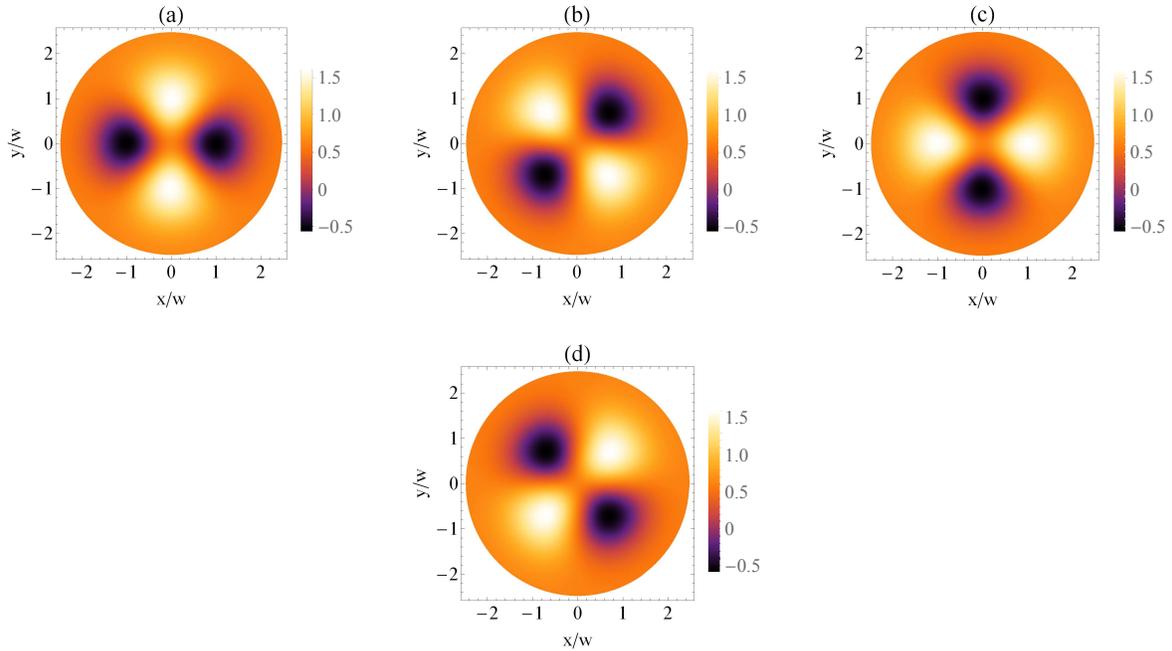}}
\caption{Spatially-dependent absorption profile of the quantum system in the presence of the plasmonic nanostructure in the case of a vortex beam with $n$=2 and $\phi=0$ in (a), $\phi=\pi/2$ in (b), $\phi=\pi$ in (c), and $\phi=3\pi/2$ in (d). The legends are also shown at the right of the figures. Common parameters are chosen as $d$=0.4 $c/\omega_{p}$, $\gamma^{'}$= 0.3 $\Gamma_{0}$, $\omega_{43}$=1.5 $\Gamma_{0}$, $E^{'}_{41}/E_{31}$=3, $\tilde{\omega}$=0.632 $\omega_{p}$, $\delta$=0, and $\mathrm{w}= 0.1$ mm.}
\label{fig2}
\end{figure*}

Before presenting the numerical results, it is desirable to point out some important considerations. The Drude model is used for the dielectric function of the shell, whose electric permittivity is given by
\begin{equation}
\epsilon(\omega)= 1- \dfrac{ \omega_{p}^{2}  }{  \omega (\omega + i/\tau) }.
\end{equation}
Here, $ \omega_{p}$ is the bulk plasma frequency and $\tau$ denotes the relaxation time of the conduction-band electrons of
the metal given by $\tau^{-1}$=0.5 $\omega_{p}$. In this paper, we consider a typical value of the plasma frequency for gold as $\hbar \, \omega_{p}$=8.99 eV. Also, the length scale of the system is considered as $c/\omega_{p} \approx$ 22 nm. Moreover, the calculations are performed on the value of distance between the quantum system and the surface of the plasmonic nanostructure: $d$=0.4 $c/\omega_{p}$. Under such assumptions, the values for spontaneous decay rates of the dipole momenta would be $\Gamma_{\bot}$=4.132 $\Gamma_{0}$ and $ \Gamma_{\|}$= 0.0031 $\Gamma_{0}$, leading to $p$ $\approx$ 0.989 \cite{Paspalakis} (also referring to figure~3 in Ref.~\cite{Evangelou}). Here, what is noteworthy is that such strong quantum interference can be also realized for distance between 0.4 $c/\omega_{p}$ and $c/\omega_{p}$. Other common parameters are $\omega_{43}$=1.5 $\Gamma_{0}$ and $E^{'}_{41}/E_{31}$=3 [$E^{''}_{41}/E_{31}$=3 for the case of the aLG beam], $\gamma^{'}$= 0.3 $\Gamma_{0}$ and $\tilde{\omega}$=0.632 $\omega_{p}$. Furthermore, we assume that both fields have equal frequencies: $\varpi_{31}$=$\varpi_{41}$. Note that our results are represented in scaled quantities; positions are divided by $\mathrm{w}$ with a typical value of the beam waist of 0.1 mm (for either the conventional or aLG field). Note that in our notation, positive (negative) values in the imaginary part of the susceptibility indicate the absorption (gain) for the probe field.

In the following, we present steady-state solutions of the probe absorption profile and demonstrate how the profile would be spatially-dependent (recalling that the steady-state solution of the master equation can be attained by setting all time derivatives of equations~(\ref{equation2}) to zero); due to the spatially-dependent interaction between the quantum system and the LG field, the dynamics of the system would be spatially-dependent. This indicates that measuring the regions of spatially-dependent transparency can be possible via measuring the absorption profile of a non-vortex probe beam. Moreover, the probe absorption depends on the azimuthal angle and the OAM of the LG field carrying an optical vortex  [$\exp(i n \varphi)$], so that it can be, in principle, possible to obtain the phase information about the OAM from such a spatially-varying absorption profile. Notice that this is completely distinct from the OAM transfer: Here, the probe beam does not acquire any OAM, instead it develops some OAM features, resulting in a spatially-varying profile. In addition, the intensity of the probe beam does not go to zero, as it has no vortex. 

First, we investigate the effects of azimuthal index and the relative phase of applied fields on the probe absorption when we have applied a conventional LG beam carrying a vortex. Interestingly, it has been found that the profile can reveal the sign of the azimuthal index of applied LG beam, in addition to detecting its modulus. Also, a spatially-varying optical transparency can be formed, due to the periodic variations of the absorption, so that the spatial inhomogeneity can be well described by the expression for the medium susceptibility given in equation~(\ref{eq5}). We then continue with configurations with aLG beams and show how these beams could alter the features of the spatial inhomogeneities.

\begin{figure*}[!t]
\centerline{\includegraphics[width=0.9\linewidth]{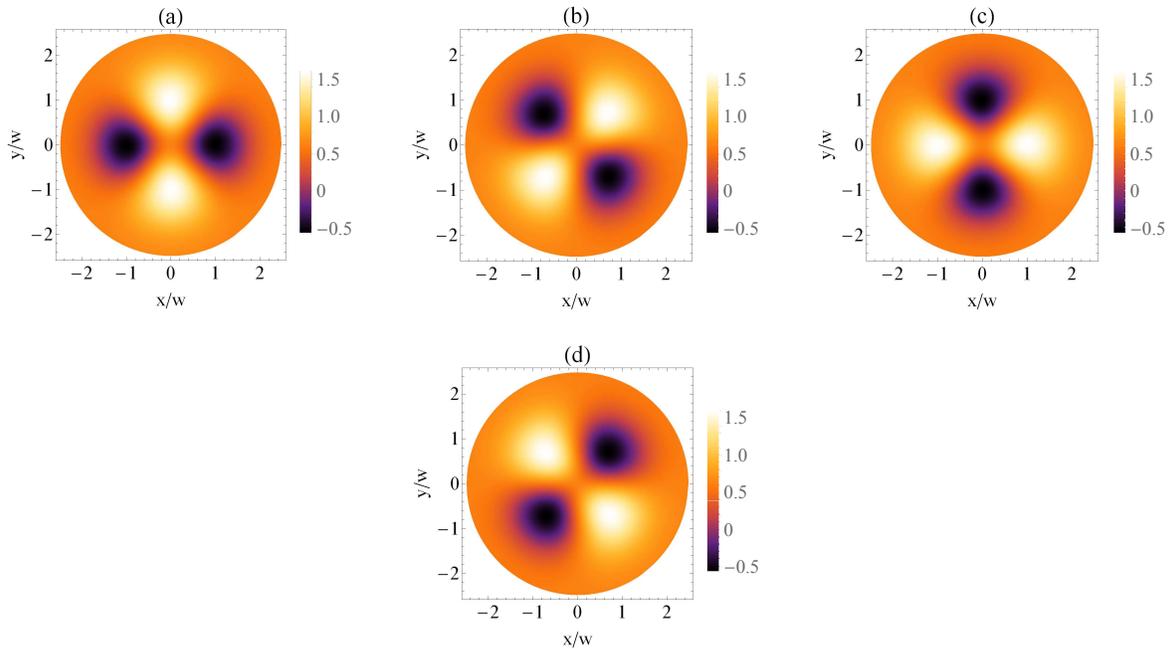}}
\caption{Spatially-dependent absorption profile of the quantum system in the presence of the plasmonic nanostructure for $n$=-2 and different relative phases of applied fields. Panels (a) to (d) correspond to $\phi=0$, $\phi=\pi/2$, $\phi=\pi$, and $\phi=3\pi/2$, respectively. Other parameters are the same as those in figure~\ref{fig2}. }
\label{fig3}
\end{figure*}

We start with studying the dependence of the spatially-varying absorption profile of the quantum system on the relative phase of applied fields. Figures~\ref{fig2}(a)-(d) show the numerical results of the probe absorption versus normalized positions ($x/\mathrm{w}$, $y/\mathrm{w}$) for a conventional LG field with azimuthal index of $n$=2 and for different relative phases of applied fields. Other parameters are chosen as $d$=0.4 $c/\omega_{p}$, $\gamma^{'}$=0.3 $\Gamma_{0}$, $\omega_{43}$=1.5 $\Gamma_{0}$, $E^{'}_{41}/E_{31}$=3, $\tilde{\omega}$=0.632 $\omega_{p}$, $\delta$=0, and $\mathrm{w}$=0.1 mm. As can be seen from the legends, the black areas indicate regions of gain in the spectra, while white areas represent positions
of large absorption. The azimuthal phase-dependent characteristics of the system can be well understand by considering the Rabi frequency of the vortex field as $ g_{41}$=$u(r)\,\exp(i n \varphi) $, with the definition of a real function as $u(r)$=$g^{'} \, (\sqrt{2} \, r/\mathrm{w})^{ \vert n \vert}\, \exp ( - r^2/ \mathrm{w}^{2})$. The susceptibility in the presence of such an inhomogeneous beam, would be proportional to -$i \,u(r)\,\exp(i n \varphi + i \phi)$. For the special case of $\phi$=0, for instance, the absorption is proportional to -$\cos(n \varphi)$: As a result of such angular dependency, it is expected that the probe absorption varies cosinusoidally with the periodicity of $n$ so that the number of appeared lobes would be equal to $2n$; therefore, an unknown vorticity of the LG beam can be easily recognized by counting the number of white (black) lobes. Noting that, identifying the OAM carried by an arbitrary light beam is of paramount importance in its possible applications in trapping, communication, and information processing.

Furthermore, the periodic variations of the absorption initiate the spatial transparency window at specific angular positions, hence, measuring the regions of spatially-varying optical transparency can be achieved by measuring the linear absorption of probe field. Such expected pattern is shown in figure~\ref{fig2}(a), showing $n$-fold symmetry. Notice that the cylindrical transverse mode pattern of LG$_{0}^{2}$ have four angularly-distributed lobes, hence this structure is mapped into the absorption profile. It is here worth mentioning that a similar trend can be also found for any value of the spontaneous decay rate ($\gamma^{'}$) and also distance between the quantum system and the surface of the plasmonic nanostructure ($d$), but with different contrasts: In fact, the moderate values of such parameters does not poses an obstacle for the formation of high-contrast patterns in the spatially-varying absorption profile. More detailedly, lower decay rates create periodic profiles with higher contrast, as compared with larger rates. In addition, the distance affects the contrast of the appeared pattern; the larger distance, the lower-contrast spatial structure.

A similar trend is also found for other values for the relative phase of applied fields, however, patterns in the absorption profile undergo an anti-clockwise rotation. Figure~\ref{fig2}(b) depicts the spatially-dependent absorption profile of the quantum system for $\phi$=$\pi/2$; here, the absorption would be proportional to a sine function [-$\sin (n \varphi)$], and consequently, the appeared pattern is expected to be rotated by -45$\degree$. Also, similar to the previous case, the absorption profile display expected $2n$ lobes. More importantly, due to the appearance of the sine function, the absorption profile can be also used as a detector of the sign of the azimuthal index; in other words, a 90$\degree$ rotation of patterns in the absorption profile is occurred by reversing the sign of the azimuthal index (this point will be explained in detail below, with reference to the concept just mentioned). 

Figures~\ref{fig2}(c) and \ref{fig2}(d) show the spatially-dependent absorption profiles for $\phi=\pi$ and $\phi=3\pi/2$, respectively. For the former case and constant azimuthal index ($n$=2), the absorption
would be proportional to $\cos(n \varphi)$ and the pattern is rotated by -90$\degree$, with respect to that in figure~\ref{fig2}(a); this means that the peaks in the absorption profile switch to dips and vice-versa, in such a way that the whole pattern remains unchanged, but with an anti-clockwise rotation. Similarly, another rotation of the pattern occurs for the case of $\phi=3\pi/2$, since the absorption contains a term of the form $\sin (n \varphi)$ [see figure~\ref{fig2}(d)]. Needless to say, the pattern will be fully rotated back to its initial condition, for the case of $\phi$=2$\pi$. Here, similar to the aforementioned cases, the number of lobes is twice the modulus of the azimuthal index associated with the LG field. 

\begin{figure*}[!t]
\centerline{\includegraphics[width=0.9\linewidth]{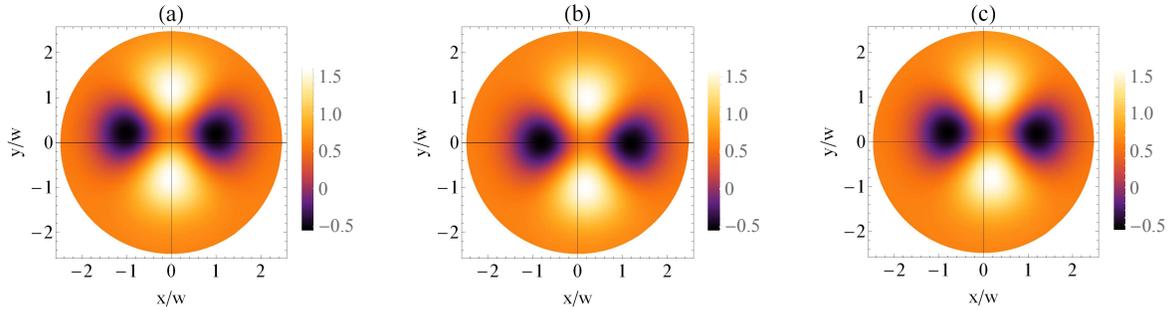}}
\caption{Spatially-dependent absorption profile of the quantum system in the presence of the plasmonic nanostructure in the case of an aLG beam with $\delta x$=0, $\delta y$=0.2 $\mathrm{w}$ in (a) $\delta x$=0.2 $\mathrm{w}$, $\delta y$=0 in (b) and $\delta x$=$\delta y$=0.2 $\mathrm{w}$ in (c). Other parameters are the same as those in figure~\ref{fig2}(a). The axes have been added in order to more clearly illustrate the displacement of the center of the patterns.}
\label{fig4}
\end{figure*}

In the following, we will show how the absorption profile can reveal the sign of azimuthal indices of the LG field. As mentioned before, the probe absorption of the system for the case of an LG field, \textit{i.e}., $ g_{41}$=$u(r)\,\exp(i n \varphi) $, and for the relative phase of zero $\phi$=0, contains a term proportional to a cosine function. As a result of such dependency, the profile do not provide any information about the sign of OAM. This fact is deduced from figure~\ref{fig3}(a), which shows the spatially-dependent absorption profile, when the sign of the azimuthal index of the LG field is flipped (with other parameters fixed); the pattern for negative index is the
same as that for positive one [compare figure~\ref{fig2}(a) and figure~\ref{fig3}(a)]. What is also worth mentioning is that the absorption profile display four-fold lobes, satisfying the 2$n$ cosinusoidal absorption profile predicted from equation~(\ref{eq5}).

However, OAM with index $n$ and -$n$ can be distinguished in our suggested scheme, by choosing proper values for the relative phase of applied fields. Figure~\ref{fig3}(b) depicts the spatially-dependent absorption profile of the system for the case of $n$=-2 an $\phi$=$\pi/2$. One can see that changing the relative phase from zero to $\pi/2$ causes a rotation of the profile by 45$\degree$. Additionally, the rotated pattern indicates that the LG field carries an opposite index [compare figures~\ref{fig2}(b) and \ref{fig3}(b)]. On the other hand, figure~\ref{fig3}(c) shows the absorption profile for the same parameters as those of figure~\ref{fig2}(c) and $n$=-2. As can be deduced from the dependency of the probe absorption on the azimuthal index [$\cos(n \varphi)$], the sign of the azimuthal index can not be distinguished by such patterns. From equation~(\ref{eq5}), a simple derivation shows that sorting out negative and positive azimuthal indices can not be possible for $\phi$=$l \pi$, with $l$ as an integer. The spatially-dependent absorption profile for $\phi=3\pi/2$ is shown in figure~\ref{fig3}(d), displaying expected trend and rotation: Having fourfold symmetry and producing a anti-clockwise rotation of 45$\degree$, with respect to that in figure~\ref{fig3}(a). In addition, the profile properly reflects the sign of OAM, as it is rotated by a angle of 90$\degree$, compared to the corresponding plot for the positive index shown in figure~\ref{fig2}(d). From two figures~\ref{fig2} and \ref{fig3} it appears that the rotating direction of the generated pattern in the absorption profile is related to the sign of the index associated with the LG beam: The profile for a positive (negative) azimuthal index undergoes an anti-clockwise (clockwise) rotation, by increasing the relative phase from zero to 2$\pi$. It is imperative to mention here that, in one of our recent works \cite{kazemi9}, we introduced the rotation of spatially-dependent absorption profiles, as a detector of the sign of the azimuthal indices of applied LG beams. In that case, a closed-loop interaction scheme in a semiconductor quantum well structure was explored to create a phase-dependent medium susceptibility. 

\begin{figure*}[!t]
\centerline{\includegraphics[width=0.9\linewidth]{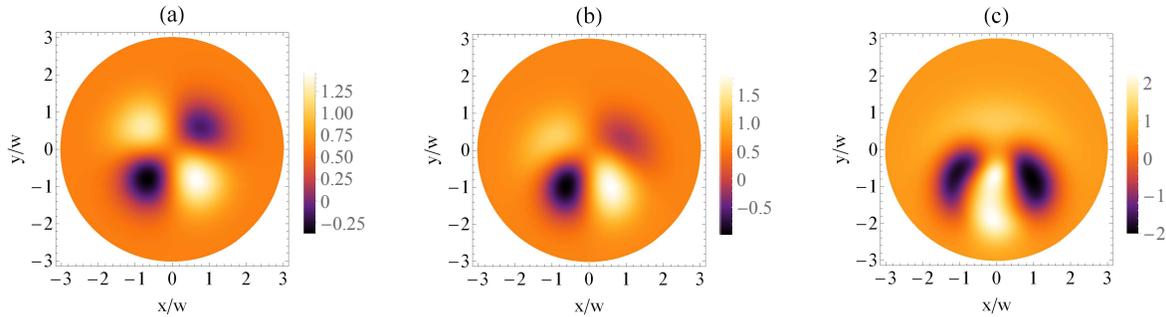}}
\caption{Spatially-dependent absorption profile of the quantum system in the presence of the plasmonic nanostructure in the case of an aLG beam with $\delta x$=-$i \delta y$= $i a \mathrm{w}$ and $a$=0.1  in (a), $a$=0.3 in (b) and $a$=0.5 in (c). Other parameters are the same as those in figure~\ref{fig2}(a).}
\label{fig5}
\end{figure*}

So far, we have demonstrated the use of number of lobes and the rotation of the pattern in spatially-dependent absorption profiles as the detectors of the azimuthal index of a conventional LG field. We then proceed to investigate the azimuthal modulation of absorption by using an aLG laser field with a zero radial index, which can be generated by using a liquid-crystal spatial light modulator \cite{Kovalev,Kovalev2}. Starting with a real shift(s), we will show how the center of appeared patterns in the spatially-dependent absorption profile can be changed, but its shape remains unchanged; hence, by counting the number of bright (dark) areas, an unknown LG mode can be easily recognized. Also, the positions of optical transparency are displaced by the real shift so that the formation of the transparency window at a desired position can be accomplished by tuning such real shifts. We then continue to extend such selective spatial transparency mechanism to asymmetric spatially-structured windows in arbitrary positions, by introducing another set of shifts and show how the structure of the absorption profile can rotate, displaced or distorted, depending on the shifts. We should reiterate that the following three cases of aLG beams with $m$=0 and $n$=2 are presented as examples of such modulation; however, this treatment can be applied to any other LG beams with non-zero radial indices. 
%Furthermore, one can implement the mechanism, in a similar way, in other closed-loop system such as atomic ones, since the shape of the LG beam is the main reason behind the results.
%and subsequently, spatially-structured optical transparency 

First, we consider the case in which one or both shifts are real; figures~\ref{fig4}(a)-\ref{fig4}(c) show spatially-dependent absorption profiles for the aLG field with real shift(s) as $(\delta x, \delta y)$=$(0,0.2 \mathrm{w})$, $(0.2 \mathrm{w},0)$ and $(0.2 \mathrm{w},0.2 \mathrm{w})$, respectively. Also, Rabi frequency of the LG field is chosen as $g^{''}_{41}$=3 $g_{31}$ and other parameters are kept at the values given in figure~\ref{fig2}(a). Moreover, the axes have been added to more clearly demonstrate the displacement of center of patterns. As seen from this figure, the absorption profile bears the same four-fold patterns as in the unshifted LG field case, however, the center of the pattern is changed as compared to the later. Accordingly, the shifted pattern would influence the spatially-structured optical transparency. Notice that the patterns have the same orientation as the conventional LG beams shown in figures~\ref{fig2}(a) and \ref{fig3}(a).

As is well known, a doughnut profile has a central intensity minimum in the origin of the coordinates, with a spherical intensity maximum, whereas the maximum and minimum intensity points of the aLG beams can be displaced. Calculations show that the central intensity minimum for such asymmetric beams is observed at the points \cite{Kovalev}
\begin{equation}
(x_{min},y_{min})=(\mathrm{Re}\,[\delta x]-\mathrm{Im}\,[\delta y], \,\mathrm{Re}\,[\delta y]+\mathrm{Im}\,[\delta x]).
\label{eq11} 
\end{equation}
Similarly, the center of appeared patterns in the spatially-dependent absorption profile would be displaced by the given vector: The center of patterns in figures~\ref{fig4}(a)-\ref{fig4}(c), respectively, is located in (0, 0.2$\mathrm{w}$), (0.2$\mathrm{w}$, 0) and (0.2$\mathrm{w}$, 0.2$\mathrm{w}$). Also notice that the profile assures that the LG field carries OAM with units of $n$, as it is evident from this figure. Worth to note, the power of aLG modes remains the same as that of the conventional LG ones, as if beams are shifted by a real distance, hence there is no need to a normalization constant in the beam amplitude. 
%Similarly, the center of appeared patterns in the spatially-dependent absorption profile would be displaced by the given vector. This is exactly what we expect, because information carried by the LG field should be converted to the absorption profile of the system.

Then, we consider the case of purely-imaginary magnitude $\delta x\, \delta y$, where the shifts have the same absolute value ($ \delta x$=-$i \delta y$). As we know from the work of Kovalev \textit{et al}, complex-valued shifts always result in an increased power of aLG beams, so a normalization constant must be included in the definition of the beam amplitude in order to keep the power of profiles (1) and (2) unchanged; \textit{i.e.}, $\mathcal{A}$= $[\, \exp(\dfrac{2 D_{0}^{2}}{\mathrm{w}^2})\,  L_{m+n}^{0} (-\dfrac{2 D_{0}^{2}}{\mathrm{w}^2}) \,  L_{m}^{0} (-\dfrac{2 D_{0}^{2}}{\mathrm{w}^2}) \, ]^{-1/2}$,  with $D_{0}$=$\sqrt{(\mathrm{Im}\,[\delta x])^2+(\mathrm{Im}\,[\delta y])^2}$ \cite{Kovalev}. Generally speaking, aLG beams can be displaced, in addition to their possible distortions, as only one shift along one coordinate is imaginary and real along the other; however, we assume here the following values for the shifts, to retain the center of the vorticity at the axis of the beam: $ \delta x$=-$i \delta y$=$i a \mathrm{w}$. 

\begin{figure*}[!t]
\centerline{\includegraphics[width=0.9\linewidth]{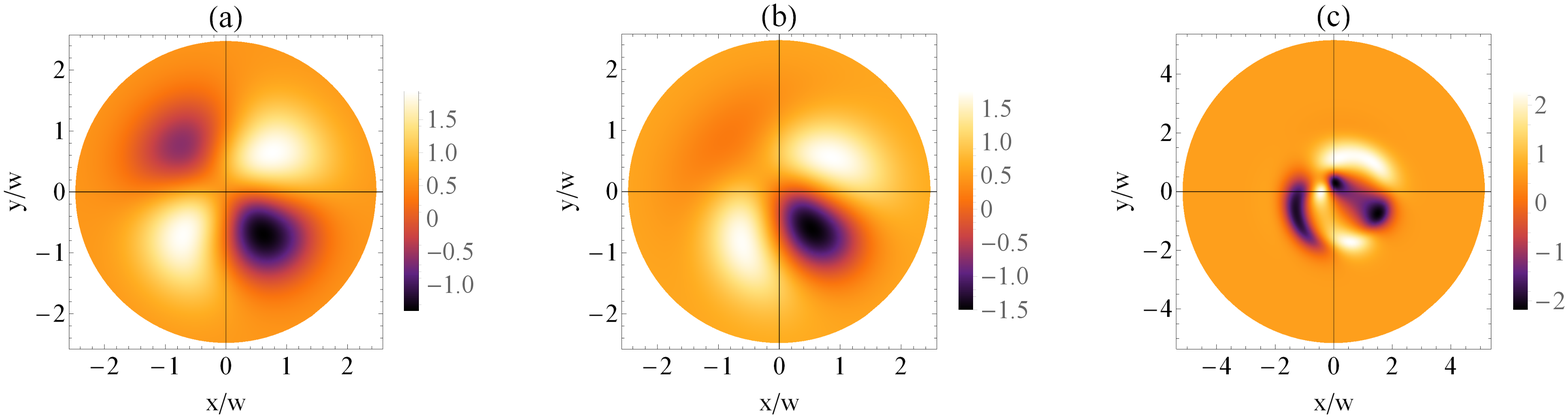}}
\caption{Spatially-dependent absorption profile of the quantum system in the presence of the plasmonic nanostructure in the case of the aLG beam with $\delta x$=$ \delta y$= $i a \mathrm{w}$ and $a$=0.1 in (a), $a$=0.3 in (b), and $a$=0.5 in (c). Other parameters are the same as those in figure~\ref{fig2}(a). The axes have been added in order to more clearly illustrate the displacement of the center of the patterns.}
\label{fig6}
\end{figure*} 

Spatially-dependent absorption profiles for the aLG fields with three different asymmetry parameters as $a$=0.1, $a$=0.3 and $a$=0.5, are shown in figures~\ref{fig5}(a)-\ref{fig5}(c), respectively. Other parameters are the same as those in figure~\ref{fig2}(a). As can be seen form this figure, the patterns appeared in the absorption profile become more and more asymmetric by increasing the asymmetry parameter ($a$). In the case of a small shift, the profile has four angularly-distribution lobes with almost equal energy for each lobe, as shown in figure~\ref{fig5}(a). While, the energy of the spots in the absorption profile with $a$=0.3 becomes unequal in such a way that two upper spots, \textit{i.e.}, those in the first and second quadrants, are slightly diminished. By further increasing the asymmetry parameter, the beam's asymmetry tries to confine the spots along the angle of -90$\degree$ [see figure~\ref{fig5}(c)]; indeed, such an LG beam encodes the information on to the probe profile around specific positions in third and fourth quadrants. Notice that the maximum value of the legend in this figure is also increased comparing with the previous case, mainly due to an increase of the power. Another remarks to be made about this figure is that such asymmetric beams induce spatial inhomogeneity in the probe absorption, and consequently, asymmetric transparency windows can be formed at specific positions. 

Such behavior can be explained in terms of the shape of the aLG field. In the case of only one imaginary shift, the transverse pattern of the asymmetric beam with small asymmetry parameter remains almost unchanged: a four petal-like structure, but with slightly asymmetric pattern. Hence, one can expect that such asymmetric structures will be imprinted to the absorption profile of the system. On the other hand, the transverse pattern tends to have more asymmetric spots and it gets concentrated near a specific region, with the increasing values of shifts. Moreover, asymmetries increase the peak intensity of the beam- which is reflected in the maximum value of the legends in the plots of the transverse pattern- similar to the one shown here for the probe absorption profile. About the rotation of the pattern, one can follow the argument in Ref. \cite{Kovalev}; the appeared light crescent in the intensity patterns of aLG beams, would be rotated by the angle of $\Phi$=$\arctan(-\mathrm{Im}\,[\delta x]/\mathrm{Im}\,[\delta y])$, which would be -$\pi/2$ for our chosen shifts. As the intensity crescent is found for large transverse shift, the asymmetry causes the pattern of absorption profiles to start rotating about the axis, and finally, to have a redistributed structure along such given angle. 
%Noting that the center of the whole patterns remains at the origin of the coordinates, as expected.

Finally, we analyze the modulation of absorption for the case in which both shifts are purely imaginary $\delta x$=$ \delta y$=$i a \mathrm{w}$. It seems reasonable to expect a similar behavior for the present case, but with increased asymmetry. Figures~\ref{fig6}(a)-\ref{fig6}(c) depict the spatially-dependent absorption profiles for the aLG field with asymmetry parameters as $a$=0.1, $a$=0.3 and $a$=0.5, respectively. Other parameters are kept the same as in figure~\ref{fig2}(a). Similar to the previous case, asymmetric absorption patterns are found, although two major differences are apparent. 1) As both shifts are imaginary, the profile exhibits pronounced distortion relative to the previous case with only one complex-valued shift. 2) As the central intensity minimum of the aLG field is found at phase-singularity points $(x_{min},y_{min})$=(-$a \mathrm{w}$, $a \mathrm{w}$), the center of the patterns in the absorption profile has now changed. Furthermore, the legends confirm that the peaks with larger height (and dips with larger depth) are found by increasing the magnitude of the shifts (recalling that the asymmetry increases the peak intensity of the aLG beam).

It is evident from figure~\ref{fig6}(a) that the aLG beam with $a$=0.1  produces a 4-lobed diagonal pattern with a near-axially symmetrical structure, similar to previous cases for the unshifted beam; however, the energy of the lobes has changed and is not equal. By further increasing the shifts, the absorption profile, shown in figure~\ref{fig6}(b), reveals an inhomogeneous pattern at -$\pi/4$ degree with vanishing the spot in second quadrant. Figure~\ref{fig6}(c), finally, shows the profile with large shifts of $i$0.5$\mathrm{w}$, resulting again in 2$n$ absorption lobes. Noting that, the absorption profile of the system is again reminiscent of the petal-mode patterns in transverse profile of corresponding asymmetric beams. Another prominent point to be noted about figure~\ref{fig6} is the orientation of the appeared lobes in the absorption profile of the system; as can be seen from this figure, the whole structure tightly bound together near the angle of -45$\degree$, for the large purely-imaginary shifts. Indeed, this behavior resembles that of a transverse pattern of an aLG beam: By increasing shifts, the field is slightly diminished in the second quadrant and the whole pattern starts to rotate, aligning at the angle of $\Phi$=-$\pi/4$ \cite{Kovalev}.

\section{Conclusions}
In this letter, we have studied the azimuthal modulation of absorption in a four-level double-V type quantum system near a plasmonic nanostructure, which induces the phase-dependent optical effects. The considered system interacts with two weak laser fields which couple the lowest state with the two upper ones in the free-space transitions. If one of the fields has the LG profile with an optical vortex, the phase-dependent medium susceptibility results in the azimuthal variation of the system. Using a density-matrix methodology, it has been found that the absorption of the non-vortex probe depends on the azimuthal factor and the OAM of the vortex LG beam. First, we have considered situations in which the system is exposed to a conventional LG beam and have investigated the effects of the azimuthal index and the relative phase of applied fields on the absorption profile of the probe field. It has been found that the spatially-varying absorption profile can reveal the modulus of the azimuthal index associated with the LG field. Interestingly, due to the phase-dependent structure of the scheme, sorting out positive and negative modes of the LG field can be also possible. Therefore, an unknown vorticity of the LG beam can be recognized through mapping the spatially-dependent absorption profile of the non-vortex probe field. Moreover, a spatially-varying optical transparency can be formed, due to the periodic variations of the resulting absorption spectra, through which obtaining information about positions of low light transition, gain or optical transparency can be provided. 

Then, by exposing the system to the aLG mode, incorporated by complex-valued shifts to a conventional LG mode, we have demonstrated how such a beam influences the features of spatial inhomogeneities. For the case of particular aLG field with a real shift(s), the fourfold symmetry in the intensity of the shifted LG field, is mapped onto the spatial profile. However, the center of appeared patterns in the spatially-dependent absorption profile is moved to a different position, which coincides with the phase-singularity point of the beam. Also, the formation of spatially-structured windows in arbitrary positions can be achieved by tuning such real shifts. Moreover, two other configurations, characterized by either one or two imaginary shifts, have been developed to extend such selective spatial transparency mechanism to
asymmetric spatially-structured windows, allowing for optical manipulation of the spatial mode at an arbitrary position. It has been found that the structure of the absorption profile can be rotated, displaced or distorted, depending on the shifts in such a way that the asymmetric features of the aLG beam could be corroborated through measuring the absorption profile. Most prominently, the asymmetry parameter tries to confine the appeared patterns along specific angular positions. 

%\section*{Acknowledgments}
%This work is supported by the University of Zanjan (Grant No. ).


\section*{References}
\begin{thebibliography}{}
\bibitem{Harris}
Harris S E 1997 \textit{Phys. Today} \textbf{50}(7) 36-42
%S. E. , “Electromagnetically induced transparency,” Phys. Today 50(7), 36–42 (1997). 
\bibitem{Boller}
Boller K J, Imamo\v{g}lu A and Harris S E 1991 \textit{Phys. Rev. Lett.} \textbf{66}(20) 2593-2596
%K. J. Boller, A. Imamolu, and S. E. Harris, “Observation of electromagnetically induced transparency,” Phys. Rev. Lett. 66(20), 2593–2596 (1991). 
\bibitem{Kasapi}
Kasapi A, Jain M, Yin G Y and Harris S E 1995 \textit{Phys. Rev. Lett.} \textbf{74} 2447
%A. Kasapi, M. Jain, G.Y. Yin, and S.E. Harris, Phys. Rev. Lett. 74, 2447 (1995).
\bibitem{Hau}
Hau L V, Harris S E, Dutton Z and Behroozi C H 1999 \textit{Nature} \textbf{397} 594
%L.V. Hau, S.E. Harris, Z. Dutton, and C.H. Behroozi, Nature (London) 397, 594 (1999)
\bibitem{Saldana}
Wu Y, Saldana J and Zhu Y 2003 \textit{Phys. Rev. A} \textbf{67} 013811
%Wu, Y., Saldana, J., Zhu, Y.: Large enhancement of four-wave mixing by suppression of photon absorption from electromagnetically induced transparency. Phys. Rev. A 67, 013811 (2003).
\bibitem{Wang3}
Z G Wang, Zhang Z Y, Che J L, Zhang Y Z, Li C B, Zheng H B and Zhang Y P 2013 \textit{Laser Phys. Lett.} \textbf{10} 095402
%Controllable ultra-narrow fluorescence and six-wave mixing under double electromagnetically induced transparency
\bibitem{Deng}
Wu Y and Deng L 2004 \textit{Phys. Rev. Lett.} \textbf{93} 143904
%Wu, Y., Deng, L.: Ultraslow optical solitons in a cold four-state medium. Phys. Rev. Lett. 93, 143904 (2004)
\bibitem{Fleischhauer}
Fleischhauer M and Lukin M D 2000 \textit{Phys. Rev. Lett.} \textbf{84} 5094
%M. Fleischhauer and M.D. Lukin, Phys. Rev. Lett. 84, 5094 (2000)
\bibitem{Lvovsky}
Lvovsky A I, Sanders B C and Tittel W 2009 \textit{Nat. Photon.} \textbf{3} 706 
\bibitem{Klimov}
Klimov V V, Ducloy M and Letokhov V S 2001 \textit{Quantum Electron.} \textbf{31} 569 
\bibitem{Mori}
K\"{u}hn S, Mori G, Agio M and Sandoghdar V 2008 \textit{Mol. Phys.} \textbf{106} 893 
\bibitem{Yannopapas}
Yannopapas V, Paspalakis E and Vitanov N V 2009 \textit{Phys. Rev. Lett.} \textbf{103} 063602 
%V. Yannopapas, E. Paspalakis, and N. V. Vitanov, Phys. Rev. Lett. 103, 063602 (2009).
\bibitem{Padgett}
Padgett M J 2017 \textit{Opt. Express} \textbf{25} 11265 
%M. J. Padgett, Opt. Express 25, 11265 (2017).
\bibitem{Allen}
Allen L, Beijersbergen M W, Spreeuw R J C and Woerdman J P 1992 \textit{Phys. Rev. A} \textbf{45} 8185
%L. Allen, M. W. Beijersbergen, R. J. C. Spreeuw, and J. P. Woerdman, “Orbital angular-momentum of light and the transformation of Laguerre-Gaussian laser modes,” Phys Rev A 45, 8185-8189 (1992).
\bibitem{8}
Truscott A G, Friese M E J, Heckenberg N R and Rubinsztein-Dunlop H 1999 \textit{Phys. Rev. Lett.} \textbf{82} 1438
\bibitem{Bhattacharya}
Bhattacharya M 2015 \textit{J. Opt. Soc. Am. B} \textbf{32}(5) B55-B60
%). Rotational cavity optomechanics. JOSA B, 32(5), B55-B60.
\bibitem{das}
Das B C, Bhattacharyya D and De S 2016 \textit{Chem. Phys. Lett.} \textbf{644} 212
\bibitem{9}
Kazemi S H and Mahmoudi M 2016 \textit{J. Phys. B: At. Mol. Opt. Phys.} \textbf{49} 245401
\bibitem{10}
Kazemi S H, Ghanbari S and Mahmoudi M 2017 \textit{J. Opt.} \textbf{19} 085503
\bibitem{Sabegh}
Sabegh Z A, Amiri R and Mahmoudi M 2018 \textit{Sci. Rep.} \textbf{8} 13840
%Spatially dependent atom-photon entanglement. Scientific reports, 8(1), 13840.
\bibitem{local}
Kazemi S H, Veisi M and Mahmoudi M 2019 \textit{J. Opt.} \textbf{21} 025401
%Atom localization using Laguerre-Gaussian beams. Journal of Optics 21 025401
\bibitem{15}
Molina-Terriza G, Torres J P and Torner L 2001 \textit{Phys. Rev. Lett.} \textbf{88} 013601
\bibitem{16}
Veissier L, Nicolas A, Giner L, Maxein D, Sheremet A S, Giacobino E and Laurat J 2013 \textit{Opt. Lett.} \textbf{38} 712
\bibitem{17}
Zhang Y, Wang X and Zhang Y Z 2018 \textit{Laser Phys. Lett.} \textbf{15} 075402
\bibitem{3}
Mair A, Vaziri A, Weihs G and Zeilinger A 2001 \textit{Nature} \textbf{412} 313
% A.Mair, A. Vaziri, G.Weihs, and A. Zeilinger, Nature (London) 412, 313 (2001).
\bibitem{4}
Gibson G, Courtial J, Padgett N J, Vasnetsov M, Pasko V, Barnett S M and Franke-Arnold S 2004 \textit{Opt. Express} \textbf{12} 5448
%G. Gibson, J. Courtial, M. J. Padgett, M. Vasnetsov, V. Pasko, S. M. Barnett, and S. Franke-Arnold, Opt. Express 12, 5448 (2004).
\bibitem{5}
Groeblacher S, Jannewein T, Vaziri A, Weihs G and Zeilinger A 2006 \textit{New J. Phys.} \textbf{8} 75
%S. Groeblacher, T. Jannewein, A. Vaziri, G. Weihs, and A. Zeilinger, New J. Phys. 8, 75 (2006).
\bibitem{Kovalev}
Kovalev A A, Kotlyar V V and Porfirev A P 2016 \textit{Phys. Rev. A} \textbf{93} 063858
%. Asymmetric laguerre-gaussian beams. Physical Review A, 93(6), p.063858.
\bibitem{Kovalev2}
Kovalev A A, Kotlyar V V and Porfirev A P 2016 \textit{Opt. Lett.} \textbf{41} 2426
%Kovalev  A.A., Kotlyar, V.V. and Porfirev, A.P., 2016. Optical trapping and moving of microparticles by using asymmetrical Laguerre–Gaussian beams. Optics letters, 41(11), pp.2426-2429.
\bibitem{Subrata}
Das S, Bhowmik A, Mukherjee K and Majumder S 2019 arXiv:1907.04090
% Transfer of orbital angular momentum superposition from asymmetric Laguerre-Gaussian beam to Bose-Einstein Condensate. arXiv preprint arXiv:1907.04090.
\bibitem{Pugatch}
Pugatch R, Shuker M, Firstenberg O, Ron A and Davidson N 2007 \textit{Phys. Rev. Lett.} \textbf{98} 203601
%R. Pugatch, M. Shuker, O. Firstenberg, A. Ron, and N. Davidson, Phys. Rev. Lett. 98, 203601 (2007).
\bibitem{Han}
Han L, Cao M, Liu R, Liu H, Guo W, Wei D, Gao S, Zhang P, Gao H and Li F 2012 \textit{Eur. Lett.} \textbf{99} 34003
%L. Han, M. Cao, R. Liu, H. Liu, W. Guo, D. Wei, S. Gao, P. Zhang, H. Gao, and F. Li, “Identifying the orbital angular momentum of light based on atomic ensembles,” Eur. Lett. 99, 34003 (2012).
\bibitem{Radwell}
Radwell N, Clark T W, Piccirillo B, Barnett S M and Franke-Arnold S 2015 \textit{Phys. Rev. Lett.} \textbf{114} 123603 
%Spatially dependent electromagnetically induced transparency.
\bibitem{Sharma1}
Sharma S and Dey T N 2017 \textit{Phys. Rev. A} \textbf{96} 033811
%Phase-induced transparency-mediated structured-beam generation in a closed-loop tripod configuration
\bibitem{Hamedi}
Hamedi H R, Kudria\v{s}ov V, Ruseckas J and Juzeli\={u}nas G 2018 \textit{Opt. Express} \textbf{26} 28249
\bibitem{Sharma}
Sharma S and Dey T N 2019 \textit{J. Opt. Soc. Am. B} \textbf{36} 960-965 
\bibitem{Marino}
Marino A M, Boyer V, Pooser R C, Lett P D, Lemons K and Jones K M 2008 \textit{Phys. Rev. Lett.} \textbf{101} 093602
%A. M. Marino, V. Boyer, R. C. Pooser, P. D. Lett, K. Lemons, and K. M. Jones, Phys. Rev. Lett. 101, 093602 (2008).
\bibitem{Walker}
Walker G, Arnold A S and Franke-Arnold S 2012 \textit{Phys. Rev. Lett.} \textbf{108} 243601 
%G. Walker, A. S. Arnold, and S. Franke-Arnold, Phys. Rev. Lett. 108, 243601 (2012).
\bibitem{cao}
Cao M, Yu Y, Zhang L, Ye F, Wang Y, Wei D, Zhang P, Guo W, Zhang S, Gao H and Li F 2014 \textit{Opt. Express} \textbf{22}20177
%Demonstration of CNOT gate with Laguerre Gaussian beams via four-wave mixing in atom vapor.
\bibitem{hanle}
Anupriya J, Ram N and Pattabiraman M 2010 \textit{Phys. Rev. A} \textbf{81} 043804
\bibitem{Steiner}
Steiner R 2011 \textit{Basic laser physics. In Laser and IPL Technology in Dermatology and Aesthetic Medicine} (Berlin: Springer) 
\bibitem{gotte}
G\"{o}tte J B and Barnett S M 2012 \textit{Light beams carrying orbital angular momentum} (Cambridge: Cambridge University Press)
\bibitem{dennis}
Dennis M R, O'Holleran K and Padgett M J 2009 \textit{Singular optics: optical vortices and polarization singularities} (\textit{Progress in Optics} vol 53) ed E Wolf (Amsterdam: Elsevier)
%10.1016/S0079-6638(08)00205-9
\bibitem{Paspalakis}
Paspalakis E, Evangelou S, Yannopapas V and Terzis A F 2013 \textit{Phys. Rev. A} \textbf{88} 053832
%Phase-dependent optical effects in a four-level quantum system near a plasmonic nanostructure,
\bibitem{28}
Agarwal G S 2000 \textit{Phys. Rev. Lett.} \textbf{84} 5500 
\bibitem{29}
Kiffner M, Macovei M, Evers J and Keitel C H 2010 \textit{Vacuum-induced processes in multilevel atoms} (\textit{Progress in Optics} vol 55) ed E Wolf (Amsterdam: Elsevier)
%M. Kiffner, M. Macovei, J. Evers, andC.H.Keitel, in Progress in Optics, edited by E.Wolf, Vol. 55 (Elsevier, Amsterdam, 2010), p. 85.
\bibitem{Evangelou}
Evangelou S, Yannopapas V and Paspalakis E 2012 \textit{Phys. Rev. A} \textbf{86} 053811 
%S. Evangelou, V. Yannopapas, and E. Paspalakis, Phys. Rev. A 86, 053811 (2012).
\bibitem{kazemi9}
Kazemi S H and Mahmoudi M 2019 \textit{Laser Phys. Lett.} \textbf{16} 076001
\end{thebibliography}
\end{document}